\documentclass[useAMS,usegraphicx]{mn2e}

\title[GR effects during eclipses]{
Probing general relativistic effects during AGN X-ray eclipses
}
\author[G. Risaliti et al.]
{G. Risaliti,$^{1,2}$ E. Nardini,$^{3}$ M. Elvis,$^2$, L.~Brenneman$^2$ \&
M.~Salvati$^1$  \\
$^1$ INAF - Osservatorio Astrofisico di Arcetri, L.go E. Fermi 5,
Firenze, Italy\\
$^2$ Harvard-Smithsonian Center for Astrophysics, 60 Garden St. 
Cambridge, MA 02138 USA {E-mail: grisaliti@cfa.harvard.edu}\\
$^3$ Dipartimento di Fisica e Astronomia, Universit\`a di Firenze, Largo E.~Fermi 2, Firenze, Italy\\
}
\begin{document}

\date{Released Xxxx Xxxxx XX}

\pagerange{\pageref{firstpage}--\pageref{lastpage}} \pubyear{2002}

\maketitle

\label{firstpage}

\begin{abstract}
Long X-ray observations of bright Active Galactic Nuclei show
that X-ray eclipses, with durations from a few hours to a few days, are rather common.
This opens up a new window of opportunity in the search for signatures of relativistic
effects in AGNs: an obscuring cloud covers/uncovers different parts
of the accretion disc at different times, allowing a direct check of the expected
pattern of disc emission. In particular, the combination of gravitational redshift and
relativistic Doppler boosting should imply strong differences between the receding
and approaching parts of an inclined thin disc. At present,  these effects may be already detectable 
with a ``lucky'' {\em XMM-Newton} or {\em Suzaku} observation of a complete eclipse by a Compton-thick cloud
(a rare, but not impossible-to-see event). In the future, higher sensitivity observatories  
will be able to perform these tests easily on tens of AGNs. This will
provide a powerful and direct way to test extreme gravity, and to probe the structure 
of AGNs in the close vicinity of the central black holes.
\end{abstract}


\section{Introduction}

Hard X-ray spectroscopy of Active Galactic Nuclei may provide a 
powerful probe of relativistic effects in the strong gravity regime.
If the iron K$\alpha$ emission line (the most prominent spectral feature
in AGN spectra above 2~keV) is produced by reflection of the primary source
by the inner parts of an accretion disc, a strong deformation of its
profile is expected (Fabian et al.~1989), depending on the
disc properties (inner and outer radius of the reflecting region, emissivity as
a function of radius), and, most notably, on the black hole spin (Laor~1990, 
Brenneman \& Reynolds~2006,Dov{\v c}iak et al.~2004).


On the observational side, such profiles have been indeed observed in the
X-ray spectra of many bright AGNs, starting from the ASCA spectrum of MCG-6-30-15 (Tanaka et al.~1995).
Today, about 20 sources have X-ray spectra showing hints of a relativistically blurred iron emission line (Brenneman \& Reynolds~2006, Nandra et al.~2007, Miller~2007).

Despite these encouraging results, there is no full agreement on the
interpretation of the observed spectral features as due to relativistic effects.
In particular, a varying, multi-component, partially covering, ionized absorber can reproduce the data
without invoking reflection by the inner regions of the accretion
disc (e.g. Turner et al. 2007, Miller et al.~2009).

For this reason, it is important to find new methods, in addition to the
analysis of the line profiles, in order to demonstrate beyond doubt the presence of 
relativistic effects, and to use them to measure the
physical parameters of the central black hole. 

One such method is the analysis of the line changes in response to continuum variability.
Recently Fabian et al.~(2009) and Zoghbi et al.~(2010) investigated
the X-ray spectrum of the Narrow Line Seyfert 1 1H0707-495, showing a broad emission feature 
at 0.5-1.5~keV, with a profile nearly identical to the iron K$\alpha$ emission line. This
feature is attributed to relativistically broadened Fe L-shell emission lines. 
A time variability analysis of this feature with respect to the continuum emission 
revealed a time lag of $\sim$30~sec, strongly suggesting an origin of the iron lines
within a few gravitational radii from the black hole.

In this Letter we discuss a new method to probe relativistic effects in AGN X-ray spectra,
taking advantage of X-ray occultations which might occur during the observations. The eclipsing clouds
cover/uncover the receding/approaching parts of the disc at different times, allowing 
the measurement of the emission from each individual zone; one can then check whether the observed pattern complies with models including general (gravitational redshift and light bending) and
special (Doppler boosting) relativistic effects. 

We will describe the  method, and discuss the probability of detecting such events in present and
future observations. In particular, we will show that it is already possible to detect these effects
today with {\em XMM-Newton} and {\em Suzaku} observations, if a complete, Compton-thick
occultation occurs: this appears to be a rare event, but not an impossible one. We will then show that the proposed test may become almost routine with future 
large-area observatories, which will be able to detect the same 
effects during partial and/or Compton-thin eclipses: 
the statistics of black hole eclipses available today guarantees a high probability of success.

\begin{figure}
\includegraphics[width=8.5cm,angle=0]{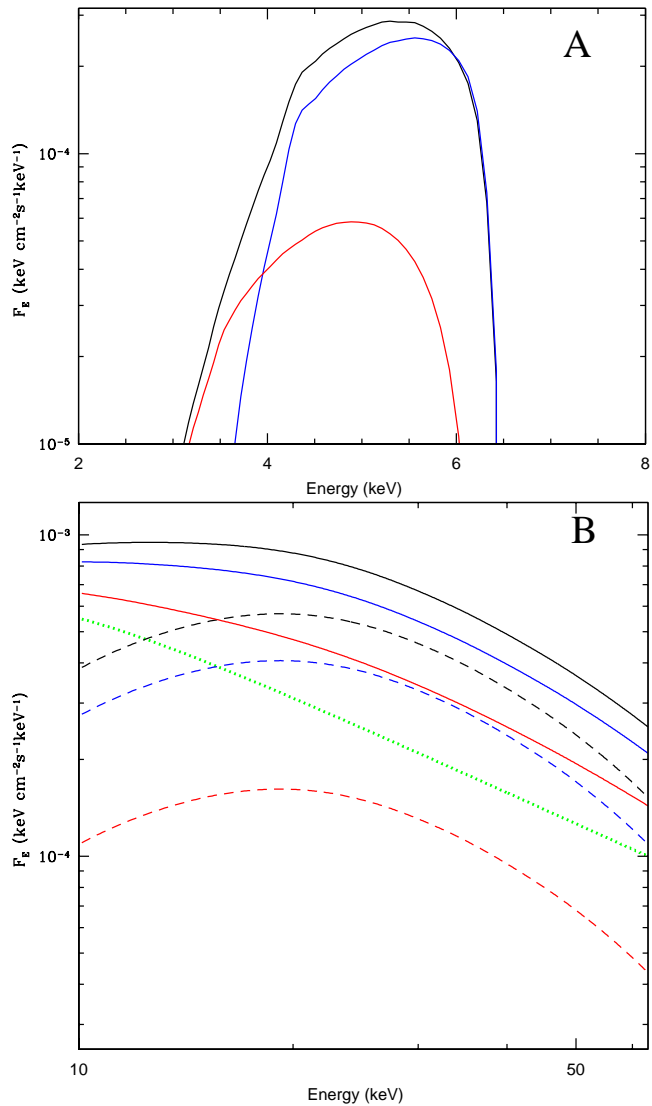}
\caption{A: Expected profiles of the iron emission line in NGC~1365
before and during an occultation due to a Compton-thick cloud. The black line shows the
total disc emission, while the blue and red lines show the contributions of the 
approaching and receding halves of the disc, respectively. The model parameters are summarized in Table~1,
and are obtained from the analysis of an {\em XMM-Newton} observation of NGC~1365 (Risaliti et al.~2009A, 2009B, Maiolino et al.~2010).
B: same, for the high-energy continuum, assuming the relativistic bending model of Dov{\v c}iak et al.~(2004).
The continuous lines have the same meaning as in the top panel. The green dotted line show the intrinsic continuum
(assumed to be constant during the eclipse). The dashed lines show the reflected components (changing at different phases of the eclipse due to the different reflection efficiencies of the two halves of the disc) }
\label{mod}
\end{figure}

\section{X-ray occultations in AGNs}

Here we review briefly the current evidence for occultations in AGNs. 
A comparison of column density measurements showed that cold absorption variability in obscured (N$_H>10^{22}$~cm$^{-2}$)
AGNs is almost
ubiquitous (Risaliti et al.~2002). In the last few years, 
several cases have been discussed where such variations occur in times scales
compatible with a single X-ray observation: the best studied case is 
NGC~1365 (Risaliti et al.~2005, 2007, 2009A, 2009B, Brenneman et al.~2011, in prep.), where several occultations have been measured,
both by Compton-thick (N$_H>$10$^{24}$~cm$^{-2}$) and Compton-thin (N$_H\sim$1-5$\times$10$^{23}$~cm$^{-2}$) clouds. For this source, the high quality of the available data allowed a detailed
analysis of the structure of the X-ray absorber.
The other well studied cases are 
NGC~4388 (Elvis et al.~2004), NGC~4151 (Puccetti et al.~2007), NGC~7582 (Bianchi et al.~2009), Mrk~766 (Risaliti et al.~2011), UGC~4203 (Risaliti et al. 2010).
A few more events have been found by our group in archival {\em XMM-Newton} observations (NGC~4395, MCG-6-30-15), and
will be discussed in forthcoming papers. 
An updated summary of the measured N$_H$ variations is reported in Risaliti~(2009). 

A complete discussion of the consequences of these observations is beyond the purpose of this Letter,
and will be presented elsewhere. In the following we only point out 
the aspects relevant  for the present work:\\
1) Eclipses on time scales of hours are {\em common} among AGNs. The above-mentioned
sources are among the 20-30 intrinsically brightest AGNs in X-ray (as found, for example, in
the {\em Swift}-BAT Catalog, Cusumano et al.~2009). 
Even if the sample 
is not yet suited for a quantitative analysis, it is large enough to conclude 
that eclipses are not rare events, in the sense
that if several bright AGNs are observed for at least a few 10$^4$~s each, we have a high probability of finding 
a few eclipsing events. \\
2) The occultation times, from a few hours to a few days, and the measured column densities, put strong constraints on the size of the X-ray source and on the size and distance of the absorber. While we refer to our papers on single eclipses for a more detailed treatment of this point, here we summarize the main argument: \\
- the limits on the ionization state of the obscuring clouds, and the estimate of the ionizing luminosity from the intrinsic X-ray flux, provide a relation between the density of the cloud (i.e. its linear size, since the column density is measured) and its distance from the center. \\
- the assumption of Keplerian velocity, and the requirement that the size of the cloud is of the same order of the size of the X-ray source (in order to obtain complete eclipses, but without long intervals with complete obscuration, in agreement with the observations) provide a second relation between the distance and size of the cloud/source.\\
The above conditions require that 
the X-ray source has a size no larger than a few gravitational radii, estimated from the available black hole mass measurements (in the range 10$^6$-10$^7$~M$_\odot$ for most of the sources with measured eclipses), and that
the obscuring clouds have tangential velocities
of several 10$^3$~km~s$^{-1}$, and densities in the range 10$^{10}$-10$^{11}$~cm$^{-3}$.
These are the typical values for Broad Emission Line clouds, which are therefore presumably one and the same with
the X-ray eclipsing clouds.\\
\section{Disc tomography by means of X-ray eclipses}

The central idea in this work is that it is possible to perform ``tomography'' experiments 
using X-ray eclipses of AGNs.

In particular, the experiment would be relevant to test relativistic effects in the inner regions around
the supermassive black hole. Two effects are expected, on both the iron emission line, and the continuum reflected emission. In the following, 
we first describe these effects physically, and we then investigate them quantitatively by means of simulations.\\
{\bf Iron line}: The broad iron K$\alpha$ emission line is believed to be produced by reflection of the primary radiation off the inner parts of the accretion disc (Fabian et al.~1989).
To a first approximation the
profile of the line does not depend on the exact geometry of the primary X-ray emitter (usually assumed to
be a hot corona around the accretion disc), while  
it is affected by both special relativistic (Doppler shift of line frequencies, and Doppler
boosting of the observed intensity) and general relativistic (gravitational redshift and light bending) effects (Fabian et al.~1989, Laor et al.~1990, Dov{\v c}iak et al.~2004).

Available models, such as Dov{\v c}iak et al.~(2004) allow the line profile from different disc regions to be computed,
taking into account the above effects, the disc inclination and a dependence of disc emissivity with radius.
In particular, the approaching and receding halves of the disc are expected to produce two markedly different
line profiles (Fig.~1A). Usually, only the total line emission can be observed, and the contributions from different regions of
the disc cannot be separated. X-ray eclipses provide a way to overcome this difficulty.\\
{\bf Reflection continuum}. If a high-EW, relativistically broadened iron line is observed, strong relativistic signatures on the
reflection continuum are also expected. These effects can be divided in two types: (1) a modification of the overall spectral shape, and (2) an increase of the ratio between reflected and primary continuum.
The former effect is hard to detect observationally, because it would require a precise knowledge of the shape of
the primary continuum. The latter effect can instead be probed observationally. In the model proposed by Miniutti \& Fabian~(2004), if the primary emission arises from a region close to the event horizon, the fraction of the
radiation illuminating the accretion disc can be much higher than the geometrical covering factor of the disc as
seen from the X-ray source, because of gravitational bending. Moreover, the iron emission line and the continuum
reflection must be treated self consistently, because both are due to reprocessing of the same primary radiation.
In practice, if a strong iron line is detected (with a higher equivalent width than observed on average in similar AGN), the ratio between reflected and primary continuum should also be higher than the standard
reflection efficiency. This effect will be particularly strong in the 20-40~keV energy range, where the contribution of the reflected emission is expected to be higher (e.g., Magdziarz \& Zdziarski~1995). Since the enhancement of the reflected flux is mainly due to Doppler boosting, a strong asymmetry in the high-energy spectrum is expected during a Compton-thick eclipse, between the ``red'' (receding) and ``blue'' (approaching) halves of the accretion disc. The high-energy continuum variation should correlate with the relativistic line variation during a Compton-thick eclipse; both components should lag (lead) the centroid of the primary continuum eclipse (in practice, the eclipse of the 2-3~keV continuum), according to the receding (approaching) half being obscured first.

\subsection{The model}

The model chosen for the simulations is based on the {\em already observed} eclipses in our best example, NGC~1365. As briefly summarized in the previous Section, and reported extensively in several publications (Risaliti et al.~1999, 2005, 2007, 2009A, 2009B, Maiolino et al.~2010), we observed several eclipses in this source, both total and partial, and by both Compton-thin and Compton-thick clouds. In particular, eclipses by clouds with N$_H\sim3-5\times10^{23}$~cm$^{-2}$, and with covering factors C$_F\sim$50-90\% are rather common, and are seen in almost all the observations longer than a few tens of ks. The duration of the eclipses is of the order of 50~ks. Compton-thick eclipses are also common, as demonstrated by the changing status (from transmission-dominated to reflection-dominated) among different observations, and in particular by a {\em Chandra} snapshot campaign of six 15~ks long observations, where a complete Compton-thick eclipse happened in a time scale shorter than two days.
Beside the eclipses, the high-quality {\em XMM-Newton} and {\em Suzaku} observations revealed two crucial spectral features: 1) a broad emission feature, with a substantial equivalent width (EW$\sim$350~eV), and well fitted with a relativistic iron line from a maximally rotating black hole (Risaliti et al.~2009A); and (2) a high-energy continuum excess, which cannot be reproduced by a primary emission-plus-partial-covering model (see Risaliti et al.~2009B for a detailed discussion), but is instead well reproduced by a self-consistent relativistic reflection model (Walton et al.~2010). 

We used for the simulations the best fit model of {\em XMM-Newton} and {\em Suzaku} spectra of NGC~1365 outside eclipses (Risaliti et al.~2009A, Walton et al.~2010). 
We will discuss in the next Section the implications of this choice.
The model parameters are summarized in Table~1. The model profiles for the different eclipse phases are shown in Fig.~1.
\begin{table}
\centerline{\begin{tabular}{lc}
Parameter & Value \\
\hline
Line Energy & 6.68$\pm0.03$ keV \\
EW     & 340$\pm35$ eV \\
Incl. angle & 24$^{+8}_{-4}$ deg. \\
R$_{IN}$ & 2.7$\pm0.2$ R$_G$ \\
R$_{OUT}$ & 400 R$_G$  (fixed)\\
q$^a$ & 4.3$^{+0.5}_{-0.4}$ \\
\hline
\end{tabular}}
\caption{Main model parameters obtained from the best fit of the {\em XMM-Newton} observation of NGC~1365, and adopted in our simulations. The inner and outer disc radii, R$_{IN}$ and R$_{OUT}$, are in units of gravitational radii R$_G$. $^a$: exponent of the power law modeling the disc emissivity as a function of radius.}
\end{table}
 
\subsection{Simulations of the ``perfect'' eclipse}  

In the following we show simulations illustrating the line profiles as could be observed with several present and future X-ray observatories.
In all cases, we assume a duration of the eclipse of 50~ks, i.e. of the same order as 
those already observed in NGC~1365 and in other sources (Risaliti~2009). 
\begin{figure*}[h!]
\includegraphics[width=17.5cm,angle=0]{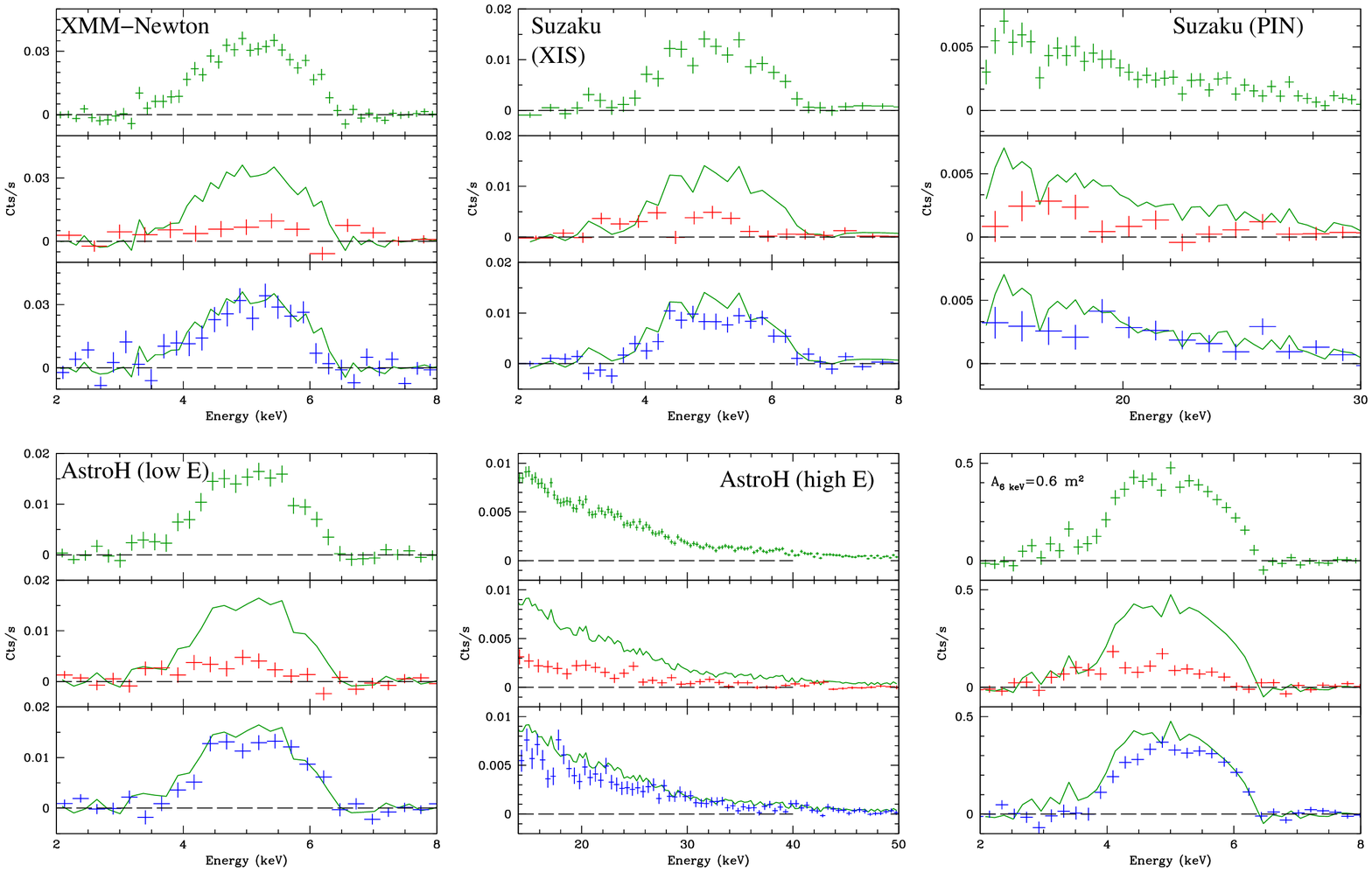}
\caption{Simulations of relativistic lines and reflection continua for present and future X-ray observatories. The adopted model is the one discussed in the text and Table~1, and shown in Fig.~1. In each plot, the top panel shows the total emission. The middle and lower panel show the emission from the approaching and receding halves of the disc, and the total emission (as a continuous line) for ease of comparison.}
\label{totfit}
\end{figure*}

We stress again the 
most interesting aspect of the simulations, and the main motivation of this work, namely that the parameters
of the model are not chosen arbitrarily, but are obtained from already observed eclipses, and are therefore expected to
correspond to easily (re-)observable events.

We start from a complete eclipse by a Compton-thick ($N_H>10^{25}$~cm$^{-2}$) cloud with sharp, linear edges. 
This is considered the ``perfect'' event because it would provide a completely separate view of the 
approaching (``blue'') and receding (``red'') halves of the accretion disc, as well as a view of the
total line emission (as long as the clouds are more or less coplanar with the disc) .

The results shown in Fig.~2 are for observations with present, or nearly completed, observatories ({\em XMM-Newton}, {\em Suzaku}, {\em Astro-H}), and for projects of possible new missions, such as {\em IXO} or {\em Athena},with an assumed effective area at 6~keV  A$_6$=0.6~m$^2$, and {\em Gravitas} or Extreme Physics Explorer (Elvis~2006), with an assumed A$_6$=2~m$^2$).

The main results are the following:
\begin{itemize}
\item The detection of the differences between the ``red'' and ``blue'' halves of the disk in the line profiles are well within the capabilities of currently available observatories. Thus, the experiment proposed here can be a strong, probably conclusive test on the relativistic effects in X-ray spectra of AGNs. Indeed, the expected asymmetry can be perhaps reproduced as a combination of partly ionized, partly covering clouds moving across the line of sight (i.e. the alternative scenario to explain the observed spectral features). However, the required peculiar timing of the line variations with respect to the primary continuum would have no physical motivation in the partial-covering scenario, while it would be required in the eclipse-plus-relativistic line scenario.
The order of the observations of the ``red'' and ``blue'' halves will also directly reveal whether the disk and the obscurer are co-rotating (``blue'' half first) or counter-rotating (``red'' half first).
\item The strongest observable effect is not the change of the line profile due to general relativistic effects, but the strong flux difference between the two halves of the disc due to Doppler boosting. 
\item The flux variations at high energies (E$>$10~keV) during a Compton-thick eclipse are quite strong, but nearly featureless. It would be impossible to distinguish between a model like the one used to simulate the data, and an intrinsic variation of a partially-covered component (with column density of the order of 10$^{24}$~cm$^{-2}$). 
As a consequence, in a realistic case the high-energy variation would provide a strong support to the relativistic scenario only if coupled with the variation of the iron line. 
\item The signal-to-noise of the variations observed by possible future instruments with high effective area at 6~keV is quite  high. These effects might be easily detected by such instruments in much shorter eclipses, or in objects where the iron line is less prominent, or the eclipse is not complete. These points will be further discussed from a general point of view in Section~4, while a special case is extensively described in the next Subsection.
\end{itemize}

\subsection{The ``likely'' eclipse}
The case discussed above demonstrates that the analysis of line profile variations is within the possibilities of current
X-ray observatories, if optimal conditions are met. 
The observational history of NGC~1365 shows that such conditions are {\em possible} (they already occurred at least once, though not during a continuous monitoring) but {\em rare} and difficult to catch (NGC~1365 has been monitored for long intervals several times, and the ``perfect'' eclipse has not been observed yet).
\begin{figure}
\includegraphics[width=8.5cm,angle=0]{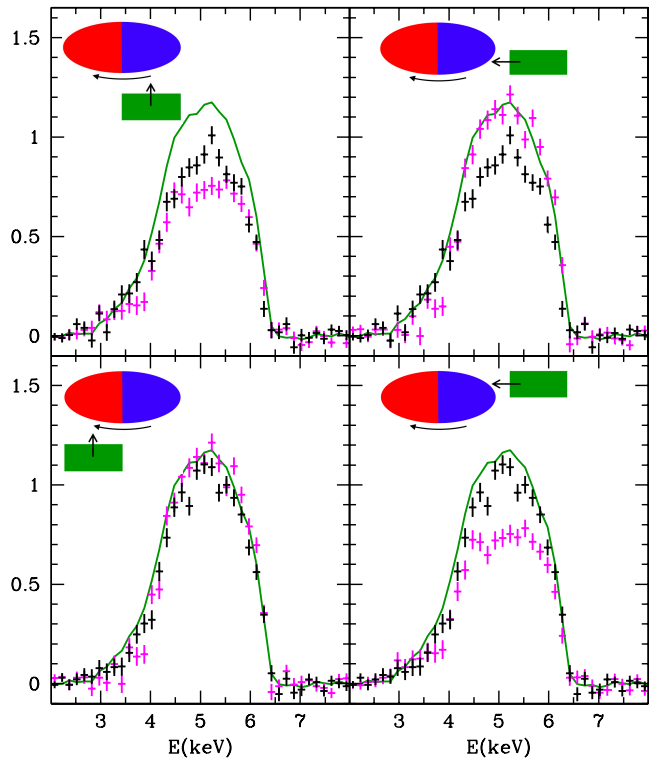}
\caption{Simulations of partial eclipses by a Compton-thin cloud (N$_H$=3.5$\times$10$^{23}$~cm$^{-2}$) observed by a high effective area observatory (2~m$^2$ at 6~keV). In each case the cloud (the dark green rectangle in each figure) covers only 25\% of the source at a time. The different profiles depend on which of the four disc quarters is covered by the cloud. In the figures, the blue and red colors indicate the approaching and receding halves of the disk with respect to the observer (whose line of sight is perpendicular to the plane of the figure). The disk bottom part, as viewed in the figure, is closer to the observer. In each panel three spectra are shown: the total emission, and the emission with the cloud covering each of the two quarters affected by the eclipse. The time sequence is color-coded (pink first, black second) }
\label{totfit}
\end{figure}

Here we discuss the possibility of performing a similar study of line profile variations during ``normal'' eclipses, such as the ones observed several times in NGC~1365 and in the other AGNs listed in the previous Sections. 
The average properties of these eclipses are the following (Maiolino et al.~2010, Risaliti et al.~2009B:\\
- Column density of the obscuring clouds of the order of a few 10$^{23}$~cm$^{-2}$;\\
- Duration of 10-15~hours;\\
- Partial covering of the X-ray source, with the covered fraction of the order of 50-80\%.\\

With these parameters, {\em XMM-Newton} and {\em Suzaku} do not provide enough counts to perform line variability studies.
This has been confirmed by our simulations, and, more importantly, by the analysis of the available observations.

Instead, simulations for large-area future observatories show that assuming the above parameters (with a conservative covering factor of only 25\%),
one would be able to detect line profile variations with high confidence. The simulations in Fig.~3 show four possible combinations depending on the direction of the eclipsing cloud with respect to the disc axis.
Clearly, we have the highest contrast between the different phases of the eclipse when the cloud covers a ``blue'' 
region of the disc, while the difference is smaller when only the ``red'' half is eclipsed.

With the exception of the unfortunate case of panel C, the proposed experiment should become routine with the next-generation observatories.

\section{Discussion}
We have shown that X-ray eclipses of AGNs can provide a new, powerful probe of the relativistic effects in the inner
parts of accretion discs around supermassive black holes. 
In particular, a complete eclipse by a Compton-thick cloud, such as the one already occurred during a snapshot {\em Chandra} campaign (and therefore, because of the gaps in the coverage, not useful for this kind of analysis) would allow us to detect the line profile variation from the ``blue'' (approaching) half of the disc to the ``red`` (receding) one with {\em XMM-Newton}.
Moreover, we have shown that ``common'' eclipses, such as the ones due to Compton-thin clouds covering only part of the X-ray source, which are easily observed in nearby bright AGNs, would be enough to perform the experiment with future high-effective area observatories. 

The success of the tomography experiment described here depends on several uncertain aspects:\\
1) The structure of the obscuring clouds may be more complex than the simple one assumed here, with column density gradients across the line of sight, and shapes different from spherical (Maiolino et al.~2010). Indeed, the simulations discussed here, which assume a perfect covering of a half disc at a given time, refer to clouds with a linear edge. This is a reasonable approximation for clouds much larger than the X-ray source, but not for spherical clouds with comparable dimensions.\\
2) The possible variations of the intrinsic continuum during the eclipse are unknown. These variations have two different origins: the intrinsic flux variability (which is present in most of the sources with observed eclipses), and the eclipse of the X-ray primary source itself (which has an unknown geometry, and therefore is covered in a different way with respect to the reflecting disc). 

As a consequence of these phenomena, we cannot assume a constant ratio between the intrinsic and reflected components. We can use the observed flux around 2-3~keV as a proxy to the intrinsic flux; still, we need to assume a constant spectral index in order to extrapolate to the high-energy region.\\
3) The geometrical and physical parameters of the black hole/disc can affect the observed spectra during the eclipse. For example, the disc inclination is obviously a fundamental parameter: we would not observe any asymmetry in a perfectly face-on disc. The black hole spin, and the inner radius of the disc are also relevant issues: the higher the reflection from the inner parts of the disc (where both the Doppler boosting and the gravitational redshift are higher) the stronger the observed asymmetry. 

All these complications/limitations have several observational consequences. In some cases (e.g. face on disk, or disk inner radius larger than a few R$_G$), the experiment would be impossible, i.e. even with a ``perfect'' eclipse, we would not be able to see any asymmetry related to relativistic effects. 

In many other cases the observations would be different from the ones described here. While it is impossible to discuss all the possible physical situations, due to the large number of free parameters and geometrical uncertainties, we make a few general comments:\\

1) The main aim of the proposed experiment is to detect an asymmetry during the different phases of the eclipse between primary and reflected emission, which would be explained {\em only} within a relativistic scenario. However, the parameters of the disc/black hole system would be better determined by the spectral analysis in the time intervals {\em without} the eclipse, where the statistics on the line and continuum emission is much larger. In other words, our ideal experiment would consist in the following steps:
\begin{itemize}
\item A spectrum with a prominent emission feature in the 4-7~keV interval, and possibly a high-energy excess, is observed. At this point, two interpretations are possible: the one based on the relativistic blurring, and the one based on variable partly ionized absorbers. 
\item An eclipse occurs, and strong time asymmetries similar to the ones discussed here are observed. These asymmetries are expected {\em only} in a relativistic blurring scenario.
\item
The black hole and disc parameters are measured based on the full, uneclipsed spectrum. Then, the spectra obtained at different phases of the eclipse are fitted by a model with the same global parameters, where part of the disc is obscured. 
\end{itemize}
Thus, it is not crucial to investigate a-priori the details of how the unknown geometrical and physical parameters would affect the spectra at different phases of the eclipse. What is really needed is a high quality spectrum before or after the eclipse, and the detection of some asymmetry during the eclipse. 
\\
2) Given the unknown behaviour of the intrinsic component, it is mandatory that the ratio between these two components is left free to vary in the spectral analysis of all the phases of the eclipse. In practice, the reflection excess is the residual of the total observed flux after one has removed the maximum possible power law, with no assumptions on its normalization or slope. Ideally, this should match the 2-3~keV flux, as already noted above. We note that this not only a ``complication'' of our analysis, but also an important source of information: we will be able to measure the relative dimensions of the intrinsic source, with respect to the reflecting disc, and possibly also its structure. For example, if the X-ray emission arises from a flat corona extending several R$_G$ along a plane parallel to the disc, we will observe a relatively long eclipse of the primary emission (of the same order of the eclipse of the reflected component). If instead the X-ray source is a compact, almost point-like source slightly above the black hole event horizon (as in the model of Miniutti \& Fabian~2004) we expect a much faster covering/uncovering.  \\
3) Some of the unknown parameters of the disc/black hole/absorber system may be obtained by the data themselves. For example, the order of appearance of the red and blue components will directly tell us whether the obscuring cloud is rotating in the same sense as the reflecting disc. 

\section{Conclusions}

We have discussed X-ray eclipses as a possible new way to probe strong relativistic effects in the X-ray emission of AGNs.
Our main conclusions are the following:\\
1) The asymmetries in the X-ray spectra at different phases of an eclipse (with respect to the eclipse center defined by the primary emission light curve) can prove the uniqueness of the interpretation based on the relativistic blurring models. \\
2) The tomography experiment described here may be already possible with present observatories ({\em XMM-Newton} and {\em Suzaku}), if a Compton-thick cloud covers completely an X-ray source with strong relativistic features in about ten hours. We know from previous observations that such an event can happen in our ``best target'' NGC~1365.\\
3) The detection of the eclipse effects would be easy with future, large-area ($>$0.5~m$^2$ at 6~keV) instruments even in partial eclipses by Compton-thin clouds, such as the ones already observed several times in X-ray bright AGNs.
As a consequence, when observatories like {\em IXO}, {\em Athena}, {\em Gravitas} or Extreme Physics Explorer,  are available, the analysis described here will become an easy and conclusive test of the nature of the relativistic nature of the features in the X-ray spectra of AGNs.
 	
The simulations presented here have been chosen based on the eclipse parameters measured in NGC~1365, and on the available models of reflection by different regions of the disc. 
An obvious developement of our work is the calculation of the expected spectra at each  phase of the eclipse, not only at ``half phases'' as available today. This will allow a more precise fit of the observed features, which will become especially important with high-area observatories. These new models and simulations will be presented in a forthcoming paper (Brenneman et al.~2011, in prep.).
 
\section*{Acknowledgements}
This work has been partly
supported by grants 
NASA NNX08AN48G.



\begin{thebibliography}{}
\bibitem[\protect\citeauthoryear{Barcons et 
al.}{2011}]{2011arXiv1102.2845B} Barcons X., et al., 2011, arXiv, 
arXiv:1102.2845
\bibitem[\protect\citeauthoryear{Bianchi et 
al.}{2009}]{2009ApJ...695..781B} Bianchi S., Piconcelli E., Chiaberge M., 
Bail{\'o}n E.~J., Matt G., Fiore F., 2009, ApJ, 695, 781
\bibitem[\protect\citeauthoryear{Brenneman 
\& Reynolds}{2006}]{2006ApJ...652.1028B} Brenneman L.~W., Reynolds C.~S., 2006, ApJ, 652, 1028
\bibitem[\protect\citeauthoryear{Cusumano et 
al.}{2010}]{2010A&A...524A..64C} Cusumano G., et al., 2010, A\&A, 524, A64 
\bibitem[\protect\citeauthoryear{Dov{\v c}iak, Karas, 
\& Yaqoob}{2004}]{2004ApJS..153..205D} Dov{\v c}iak M., Karas V., Yaqoob T., 2004, ApJS, 153, 205
\bibitem[\protect\citeauthoryear{Elvis et al.}{2004}]{2004ApJ...615L..25E} 
Elvis M., Risaliti G., Nicastro F., Miller J.~M., Fiore F., Puccetti S., 
2004, ApJ, 615, L25
\bibitem[\protect\citeauthoryear{Elvis}{2006}]{2006SPIE.6266E..20E} Elvis 
M., 2006, SPIE, 6266, 20
\bibitem[\protect\citeauthoryear{Fabian et al.}{1989}]{1989MNRAS.238..729F} 
Fabian A.~C., Rees M.~J., Stella L., White N.~E., 1989, MNRAS, 238, 729 
\bibitem[\protect\citeauthoryear{Fabian et al.}{2009}]{2009Natur.459..540F} 
Fabian A.~C., et al., 2009, Natur, 459, 540
\bibitem[\protect\citeauthoryear{Laor}{1991}]{1991ApJ...376...90L} Laor A., 
1991, ApJ, 376, 90
\bibitem[\protect\citeauthoryear{Maiolino et 
al.}{2010}]{2010A&A...517A..47M} Maiolino R., et al., 2010, A\&A, 517, A47 
\bibitem[\protect\citeauthoryear{Miller}{2007}]{2007ARA&A..45..441M} Miller J.~M., 2007, ARA\&A, 45, 441 
\bibitem[\protect\citeauthoryear{Miller, Turner, 
\& Reeves}{2009}]{2009MNRAS.399L..69M} Miller L., Turner T.~J., Reeves J.~N., 2009, MNRAS, 399, L69 
\bibitem[\protect\citeauthoryear{Miniutti 
\& Fabian}{2004}]{2004MNRAS.349.1435M} Miniutti G., Fabian A.~C., 2004, MNRAS, 349, 1435 
\bibitem[\protect\citeauthoryear{Nandra et al.}{2007}]{2007MNRAS.382..194N} 
Nandra K., O'Neill P.~M., George I.~M., Reeves J.~N., 2007, MNRAS, 382, 194
\bibitem[\protect\citeauthoryear{Puccetti et 
al.}{2007}]{2007MNRAS.377..607P} Puccetti S., Fiore F., Risaliti G., 
Capalbi M., Elvis M., Nicastro F., 2007, MNRAS, 377, 607 
\bibitem[\protect\citeauthoryear{Risaliti et 
al.}{2005}]{2005ApJ...623L..93R} Risaliti G., Elvis M., Fabbiano G., Baldi 
A., Zezas A., 2005, ApJ, 623, L93
\bibitem[\protect\citeauthoryear{Risaliti et 
al.}{2007}]{2007ApJ...659L.111R} Risaliti G., Elvis M., Fabbiano G., Baldi 
A., Zezas A., Salvati M., 2007, ApJ, 659, L111
\bibitem[\protect\citeauthoryear{Risaliti et 
al.}{2009}]{2009ApJ...696..160R} Risaliti G., et al., 2009, ApJ, 696, 160 
\bibitem[\protect\citeauthoryear{Risaliti et 
al.}{2009}]{2009MNRAS.393L...1R} Risaliti G., et al., 2009, MNRAS, 393, L1 
\bibitem[\protect\citeauthoryear{Risaliti}{2010}]{2010IAUS..267..299R} 
Risaliti G., 2010, IAUS, 267, 299
\bibitem[\protect\citeauthoryear{Risaliti et 
al.}{2010}]{2010MNRAS.406L..20R} Risaliti G., Elvis M., Bianchi S., Matt 
G., 2010, MNRAS, 406, L20
\bibitem[\protect\citeauthoryear{Risaliti et 
al.}{2011}]{2011MNRAS.410.1027R} Risaliti G., Nardini E., Salvati M., Elvis 
M., Fabbiano G., Maiolino R., Pietrini P., Torricelli-Ciamponi G., 2011, 
MNRAS, 410, 1027
\bibitem[\protect\citeauthoryear{Tanaka et al.}{1995}]{1995Natur.375..659T} 
Tanaka Y., et al., 1995, Natur, 375, 659
\bibitem[\protect\citeauthoryear{Turner et 
al.}{2007}]{2007A&A...475..121T} Turner T.~J., Miller L., Reeves J.~N., Kraemer S.~B., 2007, A\&A, 475, 121
\bibitem[\protect\citeauthoryear{Walton, Reis, 
\& Fabian}{2010}]{2010MNRAS.408..601W} Walton D.~J., Reis R.~C., Fabian A.~C., 2010, MNRAS, 408, 601
\bibitem[\protect\citeauthoryear{Zoghbi et al.}{2010}]{2010MNRAS.401.2419Z} 
Zoghbi A., Fabian A.~C., Uttley P., Miniutti G., Gallo L.~C., Reynolds 
C.~S., Miller J.~M., Ponti G., 2010, MNRAS, 401, 2419
\end{thebibliography}
\end{document}